\documentclass{article}

\usepackage{arxiv}

\usepackage[utf8]{inputenc} % allow utf-8 input
\usepackage[T1]{fontenc}    % use 8-bit T1 fonts
\usepackage[super,sort&compress,comma]{natbib} 
\usepackage{amsmath}
\usepackage[version=4]{mhchem}
\usepackage{graphicx}
\usepackage{hyperref}       % hyperlinks
\usepackage{url}            % simple URL typesetting
\usepackage{booktabs}       % professional-quality tables
\usepackage{amsfonts}       % blackboard math symbols
\usepackage{nicefrac}       % compact symbols for 1/2, etc.
\usepackage{microtype}      % microtypography
\usepackage{lipsum}
\usepackage{amssymb}
\usepackage{bm}
\usepackage{gensymb}
\usepackage[format=plain,justification=justified,singlelinecheck=false,font={stretch=1.125,small},labelfont=bf,labelsep=space]{caption}

\title{Ab initio machine learning simulation of calcium carbonate from aqueous solutions to the solid state}

\author{
  Pablo M. Piaggi \\
  CIC nanoGUNE BRTA, Tolosa Hiribidea 76, 20018 Donostia-San Sebastián, Spain \\
  Ikerbasque, Basque Foundation for Science, 48013 Bilbao, Spain \\
  \texttt{pm.piaggi@nanogune.eu}
  \And
  Julian D. Gale \\ 
  School of Molecular and Life Sciences, Curtin University, PO Box U1987, Perth, WA 6845, Australia \\
  \And
  Paolo Raiteri \\
  School of Molecular and Life Sciences, Curtin University, PO Box U1987, Perth, WA 6845, Australia \\
}

\begin{document}

\maketitle

\begin{abstract}
A first principles machine learning model has been developed aimed at studying the formation of calcium carbonate from aqueous solution using molecular dynamics simulations. 
The model, dubbed SCAN-ML, reproduces accurately the potential energy surface derived from ab initio density-functional theory within the SCAN approximation for the exchange and correlation functional.
A broad range of properties have been calculated relevant to ions in solution, solid phases, and the calcite/water interface.
Careful comparison with results from experiments and semi-empirical force fields shows that SCAN-ML provides an excellent description of this system, surpassing state-of-the-art force fields for many properties, while providing a benchmark for many quantities that are currently beyond the reach of direct ab initio molecular dynamics.
A key feature of SCAN-ML is its ability to capture chemical reactions, which reveals that calcium carbonate ion pair formation occurs predominantly via binding of calcium to bicarbonate, with the subsequent loss of a proton to water, rather than by direct association. Our model thus paves the way for the ab initio study of reactive crystallization pathways in biominerals, which are currently poorly understood.
\end{abstract}

Calcium carbonate (\ce{CaCO3}) plays a key role in a wide array of phenomena relevant for geosciences, biology, and industrial processes\cite{rieger2014formation}.
Formation of calcium carbonate provides a natural mechanism for carbon sequestration and regulation of ocean acidity, while the biomineralization of \ce{CaCO3} is central to many organisms, such as corals and molluscs, and can be regarded as a major achievement of evolution. Conversely, the crystallization of \ce{CaCO3} can also be undesirable, as it leads to scale formation in domestic and industrial systems that employ hot water, such as boilers, heat exchangers and desalination plants\cite{somerscales1990fouling}.
Given their significance, the mechanisms by which \ce{CaCO3} crystallizes have received considerable attention over the years, and despite major advances some aspects remain the subject of intense debate\cite{kimura2022possible,katsman2022mechanism,king2022solvent}.
The early stages of this process are complex for several reasons.
For one thing, \ce{CaCO3} has several polymorphs including calcite, aragonite, vaterite, and an amorphous phase, not to mention two known hydrates.
Such rich polymorphism can lead to indirect crystallization pathways, for example, the formation of an amorphous and/or less stable crystalline precursor before the appearance of the final crystalline phase\cite{nielsen2014situ}. 
Even prior to this stage, there has been considerable debate over the mechanism of nucleation and whether it follows a traditional classical approach, or as proposed more recently it follows the pre-nucleation cluster pathway that involves progressive assembly of dynamic oligomers and a liquid-liquid phase transition\cite{gebauer2008stable,gebauer2014pre}.

Molecular simulations have contributed important insights into the microscopic mechanism of crystallization of \ce{CaCO3}\cite{quigley2008free,tribello2009molecular,demichelis2011stable,raiteri2010water,henzler2018supersaturated}.
However, the reliability of the predictions of such simulations hinge upon an accurate description of the stability of polymorphs and the thermodynamics of the ions in solution.
For this reason, several models for the interatomic interactions, also called force fields, with increasing levels of accuracy, have been proposed over the years. Despite their simplicity, such models can be useful, provided they are parameterized to accurately reproduce the solubility of the solid phase in water.
The first such models were semi-empirical and based on the rigid-ion approximation, i.e., with fixed charges located at the  nuclei\cite{freeman2007new,raiteri2010derivation,armstrong2023solubility}. 
A fundamental limitation of such models is the absence of changes in charge distribution with variations in the local environment (polarization), known to be crucial to describe ions in solution\cite{zeron2019force}.
Polarizable semi-empirical models have also been developed\cite{bruneval2007molecular} and they have been shown to provide an improved description of structure and thermodynamics when parameterized against appropriate data\cite{raiteri2020ion}.
A key shortcoming of all force fields mentioned above is their inability to describe chemical reactions, i.e., bond forming and breaking.
Capturing the carbonate (\ce{CO3^{2-}}) to bicarbonate (\ce{HCO3^-}) chemical reaction, which occurs via proton transfer, is essential to describe the crystallization process of \ce{CaCO3} at near-neutral pH, conditions at which bicarbonate is the dominant species in solution\cite{huang2021uncovering}. While this can be partially captured through reactive force field models,\cite{gale2011reaxff} it remains a challenge to determine the many parameters required to accurately reproduce such energy landscapes.

An alternative to molecular dynamics based on semi-empirical models, is ab initio molecular dynamics (AIMD) driven by forces calculated on-the-fly from quantum-mechanical density-functional theory (DFT) calculations\cite{Hohenberg64,Kohn65,Car85}.
AIMD simulations are, in principle, highly accurate and naturally capture polarization, as well as reactivity.
However, to date their high computational cost has prevented their application to study the crystallization of \ce{CaCO3} beyond limited studies of ion pairing\cite{henzler2018supersaturated,raiteri2020ion}.
Recently, the use of machine learning (ML) potentials trained on DFT energies and forces has become an effective and popular path to solving this conundrum\cite{Behler07}.
Indeed, ML potentials based on DFT calculations preserve the accuracy of AIMD at a computational cost that rivals traditional force fields.

Here, we develop one such ab initio ML potential aimed at studying \ce{CaCO3} crystallization from aqueous solution.
We compute a variety of structural and thermodynamic properties using system sizes and total simulation times well out of reach of direct AIMD.
We find that the accuracy of our first principles model greatly exceeds that of earlier rigid-ion models\cite{raiteri2010derivation}, and rivals that of state-of-the-art rigid ion\cite{armstrong2023solubility} and polarizable models\cite{raiteri2020ion} that were carefully fitted to experimental properties.
Furthermore, our model is fully reactive and captures the subtle changes in free energy barriers for proton transfer during the carbonate to bicarbonate transformation, and corresponding water dissociation, as a function of the separation distance from calcium.
Finally, we show that long-range electrostatic interactions can be incorporated from first principles via the use of Wannier centroids.
Electrostatic interactions make a non-negligible contribution to the ion association free energy, and thus cannot be ignored if accurate thermodynamics are required.

\subsection*{Training of a machine learning potential}

\begin{figure}
\begin{center}
\includegraphics[width=\textwidth]{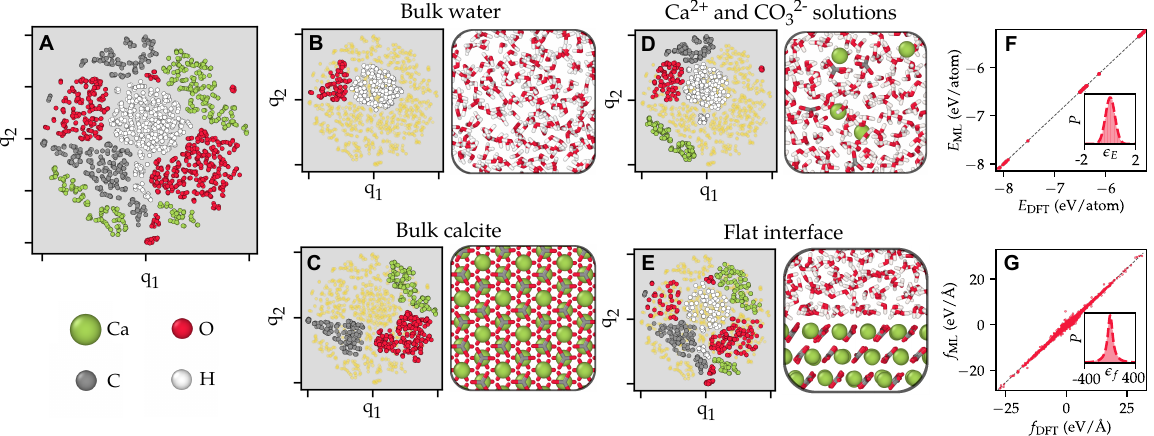}
\caption{\label{fig:Fig1} Training set and model accuracy. A) Schematic of atomic environments in the training set depicted using a two-dimensional t-SNE map based on the atomic descriptors derived from the DeePMD ML potential. $q_1$ and $q_2$ are abstract and unitless coordinates resulting from the t-SNE dimensional reduction. Environments are colored according to the chemical species of the central atom. B, C , D, and E show the subset of environments in configurations of bulk water, bulk calcite, ions in solution, and the basal surface of calcite (flat interface). The parity plots in F and G compare the energy and the forces, respectively, calculated with DFT and the ML potential. The insets in F and G show the distribution of the error in the energy ($\epsilon_E$ in meV/atom) and forces ($\epsilon_f$ in meV/\AA), respectively. }
\end{center}
\end{figure}

To construct our ML potential for \ce{CaCO3}, including in an aqueous environment, we created a training set of configurations with their corresponding energies and forces derived from DFT calculations.
For all such calculations we employed the Strongly Constrained and Appropriately Normed (SCAN) DFT exchange and correlation (XC) functional\cite{Sun15}, which has already been applied to a variety of simulations of interfaces\cite{piaggi2024first} and aqueous solutions\cite{zhang2023dissolving}.
SCAN provides an excellent balance between accuracy and cost, and the properties of water derived from this functional are well known\cite{Chen17,Piaggi21,Zhang21}.
Further details of the DFT calculations are provided in the Materials and Methods section.

The initial training set comprised liquid water data from ref.~\citenum{piaggi2024first} and configurations extracted from MD simulations using the polarizable force field of ref.~\citenum{raiteri2020ion}.
These configurations were quite diverse including ions in solution, bulk calcite, and calcite-water interfaces.
We then trained an initial ML potential using the Deep Potential or DeePMD methodology, which is an effective and widely-used framework for constructing ML potentials\cite{Zhang18,Zhang18end} (see the Materials and Methods section for further details).
Afterwards, we incorporated progressively more configurations into our training set via an active learning procedure described in detail in refs.~\citenum{piaggi2024first,zhang2019active}, which has become standard in the training of ML potentials\cite{smith2018less,podryabinkin2019accelerating,schran2021machine}.
The procedure is based on performing molecular dynamics simulations using the ML potential trained in the previous iteration, extracting configurations in which the error in the forces is high (estimated based on the deviation of four independently trained models), computing energies and forces for such configurations using DFT, incorporating them into the training set, and training a new ML potential.
This procedure was repeated for 20 iterations, after which the error in the forces was systematically below a chosen threshold.

In Figure \ref{fig:Fig1} A-E we illustrate some of the most important configurations in the final training set using the t-distributed Stochastic Neighbor Embedding (t-SNE) dimensional reduction algorithm using local descriptors derived from DeePMD as input.
The final training set spans a large variety of local environments for water, calcium, and carbonate, and is described in more detail in the Materials and Methods section.
We shall refer to the ML potential derived here as SCAN-ML.
The accuracy of the final ML potential is analyzed in Figure \ref{fig:Fig1} F-G using parity plots and error distributions for the energies and forces.
The RMSD errors in the energies and forces are 0.4 meV/atom and 101 meV/\AA, respectively.
We note that an important limitation of this model is that interactions are short-ranged and limited to 6 \AA.
In a later section, \textit{Long-range interactions via Wannier centroids}, we address this issue and analyze its impact on ion pairing thermodynamics.

\subsection*{Structural and dynamical properties}

\begin{figure}
\begin{center}
\includegraphics[width=0.6\columnwidth]{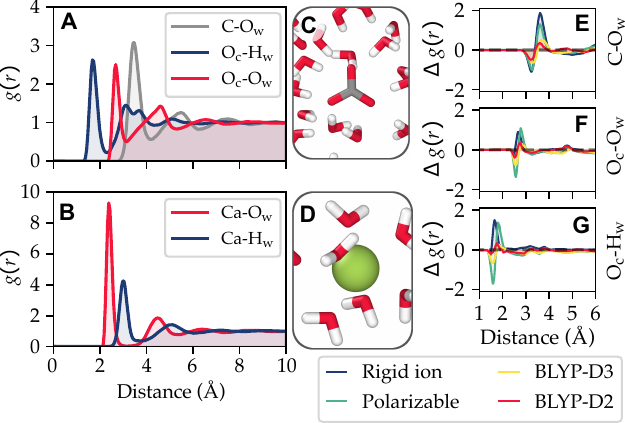}
\caption{\label{fig:Fig2} Radial distribution functions $g(r)$ between different species in systems with a single ion, either \ce{CO3^{2-}} (A) or \ce{Ca^{2+}} (B), solvated in water calculated using the SCAN-ML potential. The snapshots in panels C and D illustrate an instantaneous atomistic configuration around \ce{CO3^{2-}} and \ce{Ca^{2+}}, respectively. We also show in E, F, and G the difference $\Delta g(r)$ between the $g(r)$ for the pairs C-O$_\mathrm{w}$, O$_\mathrm{c}$-O$_\mathrm{w}$, and O$_\mathrm{c}$-H$_\mathrm{w}$, computed with SCAN-ML and state-of-the-art models reported in the literature, namely, a semi-empirical rigid ion\cite{armstrong2023solubility} potential and a polarizable\cite{raiteri2020ion} potential, and AIMD based on the BLYP-D2\cite{yadav2018structural} and BLYP-D3\cite{raiteri2020ion} functionals. O$_\mathrm{c}$, O$_\mathrm{w}$, and H$_\mathrm{w}$ refer to oxygen in carbonate, oxygen in water, and hydrogen in water, respectively.}
\end{center}
\end{figure}

We now turn to evaluating the properties of the SCAN-ML potential using molecular dynamics simulation at constant pressure (1 bar) and constant temperature (330 K).
Note that we chose a temperature somewhat higher than standard room temperature to account for the fact that the SCAN functional has many thermodynamic properties, including the melting temperature of ice Ih, shifted to higher temperatures by around 30 K\cite{Gartner20,Piaggi21}.
In Figure \ref{fig:Fig2} we characterize the structure of water around \ce{Ca^{2+}} and \ce{CO3^{2-}} ions using the radial distribution function $g(r)$ between different species. 
In the same figure, we also show the difference $\Delta g(r)$ between the $g(r)$ in our model and state-of-the-art semi-empirical models\cite{armstrong2023solubility,raiteri2020ion}.
The comparison with the $g(r)$ derived from the rigid ion model proposed in ref.~\citenum{raiteri2010derivation} is not reported because it deviates significantly with respect to the later models reported here.
This is to be expected as the rigid ion models shorten the ion-water distance to reproduce the hydration free energy without the contribution of polarization.
The rigid ion model of ref.~\citenum{armstrong2023solubility} and the polarizable model based on AMOEBA developed in ref.~\citenum{raiteri2020ion} are instead somewhat closer to our SCAN-based model.
Finally, previous $g(r)$ calculated with direct AIMD using the van der Waals corrected functionals BLYP-D2\cite{yadav2018structural} and BLYP-D3\cite{raiteri2020ion} are the closest to our SCAN-ML potential.
These results indicate that our model is significantly better for the structure of water around ions than rigid ion force fields, and on a par with direct AIMD calculations.

\begin{table}[b!]
    \centering
    \caption{Properties of single ions in aqueous solutions and calcium carbonate solid phases as determined from the SCAN-ML force field (this work), the AMOEBA polarizable force field (Ref.\ \citenum{raiteri2020ion}) and experiment\cite{ohtaki1993structure,wang1953tracer,kameda2006neutron,kigoshi1963self,yuan1974diffusion,Atkinson74,soper2013radial,kell1967precise,redfern1999high}. $D$ are diffusion coefficients (units of $10^{-5}$ cm$^2$/s), N$_{\mathrm{O}_\mathrm{w}}$ are coordination numbers with water oxygens O$_\mathrm{w}$ (unitless), $\tau$ are water residence times around Ca and the calcite basal surface, and $r_{\mathrm{A{\text -}B}}$ are the average distances between atoms A and B. The maximum distance for the calculation of coordination numbers of Ca, C, and O$_\mathrm{w}$ with O$_\mathrm{w}$ are 3.1, 4.2, and 3.3 \AA, respectively.  $a$, $b$, and $c$ are lattice constants for different solid phases. The values in parenthesis are the uncertainties in the last digit.}
    \begin{tabular}{ccccc}
         \hline
          & &  SCAN-ML & Polarizable & Experiment  \\
          & &  (This work) & (Ref.\ \citenum{raiteri2020ion}) &   \\
         \hline
         \multicolumn{5}{c}{Solutions} \\         
         \hline
         \ce{Ca} & N$_{\mathrm{O}_\mathrm{w}}$ & 6.85(5) &  7.2 &  6-10 \cite{ohtaki1993structure} \\
         & r$_{\mathrm{Ca{\text -}O}_\mathrm{w}}$ (\AA) & 2.37(5) & 2.37 &  2.33 / 2.44 \cite{ohtaki1993structure} \\
         & $D_{\mathrm{Ca}}$  & 0.9(4) & 0.9 &  0.79 \cite{wang1953tracer} \\
         & $\tau$ (ns) & 0.11(1) & 0.1 & 1.1/1.6 \cite{Atkinson74} \\
         \ce{CO3} & N$_{\mathrm{O}_\mathrm{w}}$ & 10.0(1) & 10.7   &  9.1 \cite{kameda2006neutron} \\
         & r$_{\mathrm{C{\text -}O}_\mathrm{w}}$ (\AA) & 3.45(5) & 3.55  & 3.35 \cite{kameda2006neutron}  \\
         & r$_{\mathrm{C{\text -}H}_\mathrm{w}}$ (\AA) & 2.52(5) & 2.63 & 2.68 \cite{kameda2006neutron} \\
         & $D_{\mathrm{CO}_3}$ & 1.0(6) & 0.71 &  0.8 / 0.955 \cite{kigoshi1963self,yuan1974diffusion} \\
         \ce{H2O} & N$_{\mathrm{O}_\mathrm{w}}$ & 4.47(1) & 4.6 & 4.4 \cite{soper2013radial} \\
          & r$_{\mathrm{O}_\mathrm{w}{\text -}\mathrm{O}_\mathrm{w}}$ (\AA) & 2.75(5) & 2.8 & 2.8 \cite{soper2013radial} \\
          & $D_{\mathrm{H}_2\mathrm{O}}$ & 2.4(4) &  2.36 &  2.35 \cite{kell1967precise} \\
         \hline
         \multicolumn{5}{c}{Calcite} \\
         \hline
         & $a$ (\AA) & 4.984(4) &  5.007 &  4.989 \cite{redfern1999high} \\
         & $c$ (\AA) & 17.07(2) &  17.053 &  17.061 \cite{redfern1999high} \\
         & $\tau$ (ns) &  0.80(5) & 4 \cite{brugman2020calcite} &  - \\
         \hline
         \multicolumn{5}{c}{Aragonite} \\
         \hline
         & $a$ (\AA) & 4.93(2) & - & 4.9598(5) \cite{dickens1971refinement} \\
         & $b$ (\AA) & 8.08(7) & - & 7.9641(9) \cite{dickens1971refinement} \\
         & $c$ (\AA) & 5.67(8) & - & 5.7379(6) \cite{dickens1971refinement} \\
         %& $\Delta E$ (kJ/mol) & -13(4) & - & -0.37 \\ % I am having trouble with this. DFT at 0 K gives -0.3 kJ/mol in good agreement with experiment. I think the ML potential needs to have a much higher accuracy than 1 meV/atom in order to capture this small energy difference.
         \hline
         \multicolumn{5}{c}{Monohydrocalcite} \\
         \hline
         & $a$ (\AA) & 10.6(1) & - & 10.5547(3) \cite{swainson2008structure}\\
         & $c$ (\AA) & 7.6(1) & - &  7.5644(3) \cite{swainson2008structure} \\
         \hline
    \end{tabular}
    \label{tab:Table1}
\end{table}

We also report in Table \ref{tab:Table1} a variety of properties of single ions in aqueous solution and of selected calcium carbonate crystal polymorphs.
Most properties show excellent agreement with experiment and a similar accuracy to a polarizable force field\cite{raiteri2020ion}.
Comparison with the polarizable force field is a stringent test of the accuracy of SCAN-ML, considering that it is among the best semi-empirical models for this system.

\subsection*{Thermodynamics of ion pairing}

\begin{figure}
\begin{center}
\includegraphics[width=0.6\columnwidth]{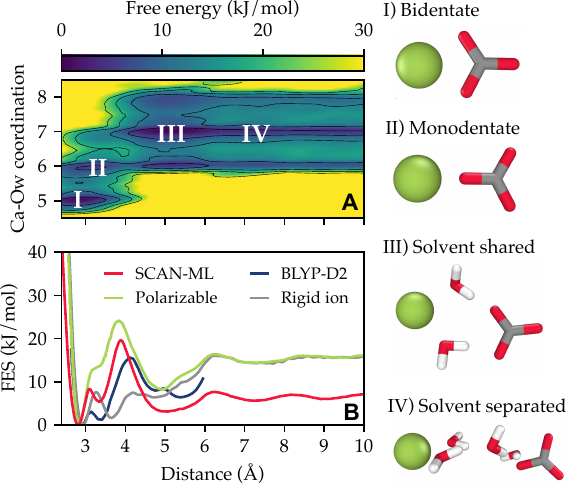}
\caption{\label{fig:Fig3} Ion pairing thermodynamics computed from averaging over four 20 ns long simulations of one ion pair in 1576 water molecules. A) Free energy as a function of the distance $d$ between \ce{Ca} and \ce{C} atoms and the Ca-O$_\mathrm{w}$ coordination number. The different states in this free energy surface are marked with white Roman numerals and representative atomic configurations are shown on the right of the figure (calcium, carbon, oxygen and hydrogen are colored green, gray, red and white, respectively). B) Free energy as a function of the distance $d$ between \ce{Ca} and \ce{C} atoms. We compare the results of our SCAN-ML model with data obtained with two semi-empirical models, namely a rigid ion\cite{armstrong2023solubility} potential and a polarizable\cite{raiteri2020ion} potential, and the result of an AIMD simulation with the BLYP-D2 functional\cite{henzler2018supersaturated}.}
\end{center}
\end{figure}

So far, we have considered properties of isolated ions in solution.
However, the interaction between ions is of fundamental importance to understanding the species present in solution, and is also the first step in the crystallization process.
We thus now investigate the thermodynamics of association of a single ion pair (\ce{Ca^{2+}} and \ce{CO3^{2-}}) as described by our model.
To this end, we performed enhanced sampling calculations to calculate the standard free energy of association by introducing a bias potential as a function of a small set of collective variables (CVs).
Here we used as a CV the distance $d$ between the \ce{Ca} and \ce{C} atoms.
We also define a second CV that was used only for analysis and is the number of water oxygen (O$_\mathrm{w}$) atoms within a sphere of radius $\sim 3$ \AA\ around \ce{Ca}, which represents a distance where the $g(r)$ for this pair reaches zero between the first and second solvation shells.
We computed the free energy as a function of $d$ and the Ca-O$_\mathrm{w}$ coordination number, as shown in Figure \ref{fig:Fig3}A.
Even if the Ca-O$_\mathrm{w}$ coordination number was not included as a CV for the construction of the bias potential, the dynamics along this degree of freedom are sufficiently fast to reconstruct the free energy shown in Figure \ref{fig:Fig3}A.
Indeed, the residence time of water around \ce{Ca} ions is $\tau\approx 0.1$ ns, which is much smaller than the total simulation time.
In Figure \ref{fig:Fig3}A we observe four states, namely, the bidentate, monodentate, solvent-shared, and solvent-separated configurations.
The Ca-O$_\mathrm{w}$ coordination number increases from 5 to 7 water molecules upon dissociation of the ion pair.
Furthermore, we show in Figure \ref{fig:Fig3}B the free energy as a function of $d$, which shows that with the SCAN-ML model the bidentate configuration is characterized by a free energy basin at around 2.9 \AA\ and is significantly more stable than the monodentate configuration, which has a shallower minimum at around 3.4 \AA.
The transition from bidentate to monodentate is accompanied by the addition of a water molecule to the solvation shell of \ce{Ca}.
There is also a sizeable free energy barrier for solvent reorganization around the ions at around 4 \AA.
This barrier is followed by the solvent-shared configuration at around 5 \AA, where the interaction between \ce{Ca^{2+}} and \ce{CO3^{2-}} is mediated by  water molecules.
Beyond 6 \AA\ the interaction of the ion pair has small yet noticeable features connected to the formation of subsequent solvation shells around ions, starting with the solvent-separated ion pair.

In Figure \ref{fig:Fig3}B we also show the ion pairing free energy curves for the rigid ion \cite{armstrong2023solubility}, and polarizable potentials \cite{raiteri2010derivation}, plus AIMD simulations with the BLYP-D2 functional\cite{henzler2018supersaturated} (for the range below 6 \AA\ for which it was calculated).
All models predict the bidentate configuration to be more stable than the monodentate counterpart.
The polarizable model does not show a defined minimum for the monodentate configuration and instead has a shoulder in that region.
Both SCAN-ML and AIMD with BLYP-D2 find a well-defined minimum, which thus seems to be the most likely scenario.
This broad consensus between the results as to the stability of the contact ion pair relative to the solvent-shared ion pair is in contrast to recent results at the revPBE-D3BJ level of theory,\cite{boyn2023camgco3} which reached the opposite conclusion. Even after applying embedded cluster-based corrections to the internal energy at the MP2 level, this study suggests that the contact and solvent-shared ion pairs are almost isoenergetic to within thermal energy. The apparent discrepancy can be explained since the AIMD for the revPBE-D3BJ study used a small simulation box that did not allow for full separation of the ion pairs from periodic images and had more limited sampling of the free energy landscape than for the other data. 
It is also important to highlight that previous rigid ion models predict monodentate to be more stable than bidentate\cite{raiteri2010derivation}, and the results shown in Figure \ref{fig:Fig3}B for the most recent rigid ion model\cite{armstrong2023solubility} are the result of decades of experience in the fitting potentials for this system.
In spite of these efforts, the ion pairing free energy of the rigid ion model overestimates the stability of the monodentate configuration and severely underestimates the barrier between the monodentate and the solvent-shared states.
Finally, there is also a marked difference in the free energy value for which the semi-empirical models and SCAN-ML reach a plateau at long distance.
This will have an impact on the ion association free energy, which we shall analyze below.

Based on the discussion above, we conclude that SCAN-ML shows a performance for ion pairing that rivals or exceeds that of the most accurate semi-empirical models in terms of describing the contact versus solvent-shared states, and provides a benchmark for the free energy landscape that is beyond the level of system size and statistical convergence that is currently accessible to ab initio MD.

\subsection*{Proton transfer during carbonate to bicarbonate transformation}

\begin{figure}
\begin{center}
\includegraphics[width=0.6\columnwidth]{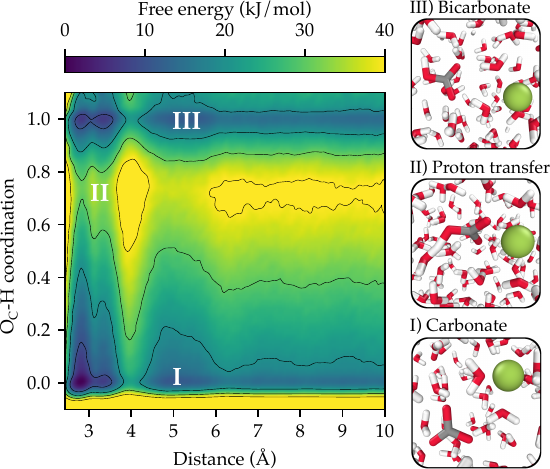}
\caption{\label{fig:Fig4} Carbonate to bicarbonate transformation. Free energy as a function of the distance $d$ between \ce{Ca} and \ce{C} atoms and the O$_\mathrm{c}$-H coordination number within a 1.3 \AA\ radius. Three relevant states are marked with white Roman numerals and the corresponding representative configurations are shown on the right.}
\end{center}
\end{figure}

As noted previously, a severe limitation of the rigid-ion and polarizable models is their inability to describe chemical reactions, i.e., bond forming and breaking.
To illustrate the reactivity of our model, we have studied the carbonate to bicarbonate transformation (accompanied by water dissociation) using enhanced sampling simulations for a total time of 50 ns; a timescale well beyond the capabilities of direct ab initio MD.
We employed two CVs, namely, the distance between \ce{Ca} and \ce{C} atoms, and the O$_\mathrm{c}$-H coordination number.
The latter CV is zero for carbonate and one for bicarbonate.
We started from a neutral system with a pair of \ce{Ca^{2+}} and \ce{CO3^{2-}} ions, and during the simulation \ce{CO3^{2-}} takes a proton from water to form \ce{HCO3^{-}} and \ce{OH^{-}}.
We observe multiple transitions between \ce{CO3^{2-}} and \ce{HCO3^{-}} mediated by proton transfer, and when \ce{HCO3^{-}} is formed the \ce{OH^{-}} does not remain bound to \ce{Ca^{2+}} but instead diffuses throughout the simulation box.
In Figure \ref{fig:Fig4} we show the free energy as a function of the two CVs described above.
The most remarkable feature of the free energy landscape is that the lowest barrier path for the carbonate to bicarbonate transformation is at an interionic distance of around 3 \AA, i.e., when the ions are tightly associated.
There is also a relevant pathway for the transformation at around 5 \AA, i.e., through the solvent-shared configuration.
Instead, when the ions have a separation above 6 \AA\ the barrier to the interconversion becomes significantly higher and exceeds 40 kJ/mol. This leads to the conclusion that the fastest and most probable pathway for calcium carbonate ion pair formation is via loss of a proton from the calcium-bicarbonate ion pair, rather than direct binding of carbonate to calcium, especially given the much higher concentration of bicarbonate under typical pH conditions.
We also show in Figure \ref{fig:Fig4} representative configurations in the free energy landscape including the proton transfer process that leads to the carbonate to bicarbonate transformation.
Another interesting feature of the free energy surface is the similar stability of the monodentate and bidentate state for the bicarbonate ion, as also found for the polarizable model\cite{raiteri2020ion}
and AIMD\cite{henzler2018supersaturated}.
Overall, the free energy landscape is quantitatively consistent with the observed pKa shift of bicarbonate on ion pairing with calcium, as well as previous indirect calculations that give a free energy shift of 9.5 kJ/mol\cite{huang2021uncovering}. However, this neglects the dependence on the relative position of the hydroxide ion when bicarbonate is present. Even though hydroxide samples a wide range of configurations throughout the simulation box, inclusion of a third CV and referencing to the standard state would be needed to accurately obtain the absolute pKa of each species. However, if we assume that the distribution of hydroxide is similar for bicarbonate and the calcium bicarbonate ion pair, which is likely if association is weak, then the relative proton transfer free energies, that yield the difference in pKa values, should be reliable.   

\subsection*{Calcite-water interface}

\begin{figure}[ht!]
\begin{center}
\includegraphics[width=0.6\columnwidth]{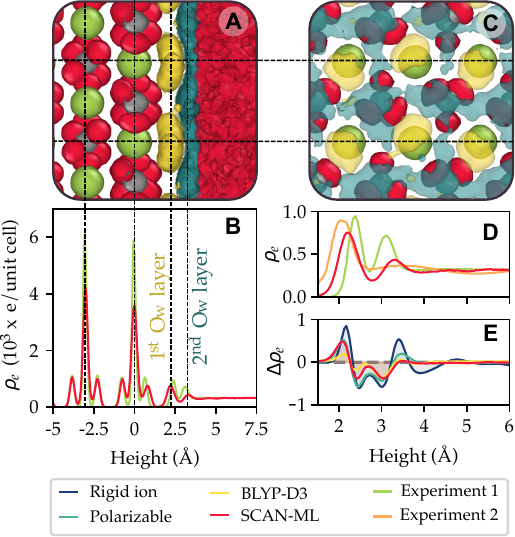}
\caption{\label{fig:Fig5} Calcite (104) surface in contact with water. A and C show lateral and top views, respectively, of the surface. Atoms are represented using isosurfaces of the atomic density using the color coding for species shown in Figure \ref{fig:Fig1}. The first and second layers of water oxygens are shown in yellow and blue, respectively. In C, only the top most \ce{Ca^{2+}} and \ce{CO3^{2-}} are shown for clarity. B and D show the electron density ($\rho_e$) as a function of the height above the surface using the last layer of \ce{Ca^{2+}} as the origin. Experiment 1 and 2 correspond to the surface x-ray diffraction results reported by Brugman et al. and Fenter et al.\cite{brugman2020calcite,fenter2013calcite}, respectively. E shows the difference in electron density ($\Delta\rho_e$) between different models reported in ref.~\citenum{brugman2020calcite} and experiment 1, which is used as a reference. The rigid ion and the polarizable model correspond to refs.~\citenum{armstrong2023solubility} and \citenum{raiteri2020ion}, respectively. $\rho_e$ is expressed in $10^3$ e/unit cell in panels A, D, and E.}
\end{center}
\end{figure}

The interface between the surfaces of calcium carbonate minerals and water are the key to understanding crystal growth processes for this system. For calcite, the most stable phase at ambient conditions, the morphology is often dominated by a single surface, namely the (104) termination. As it is possible to grow large single crystals of calcite and that the surfaces are found to be relatively clean with large terraces, the aqueous interface has been characterized experimentally, both by surface X-ray techniques\cite{fenter2013calcite,brugman2020calcite} (crystal truncation rod) and in situ atomic force microscopy\cite{RuizAgudoPutnis2018}. Consequently, this interface is an ideal benchmark for simulation models and their ability to quantitatively reproduce the observed ordered water layers. Data for the SCAN-ML model is shown in Figure \ref{fig:Fig5}.

The atomic structure of this interface is shown in panels A and C of Figure \ref{fig:Fig5}, while panels B and D show the electron density computed from a weighted sum of histograms of the nuclei density.
In Figure \ref{fig:Fig5} B we observe that the structure of calcite is in good agreement with experiment\cite{brugman2020calcite}.
Furthermore, we show in Figure \ref{fig:Fig5} D that the structure of the interfacial water shows two peaks, also in agreement with experiment, yet the position of these peaks differ somewhat.
This figure also shows that the results of the experiments reported in refs. \citenum{brugman2020calcite,fenter2013calcite} are not in good agreement with each other, and our results lie in between the results of the two experiments.
The oxygen atoms in the first water layer are highly localized, as shown in Figure \ref{fig:Fig5} A and C, and this is in good agreement with most MD models.
However, SCAN-ML predicts a somewhat diffuse second layer at variance with most classical MD models\cite{raiteri2010derivation,fenter2013calcite} which show a more structured and localized second layer.
To compare different models, in Figure \ref{fig:Fig5} we show the difference in electron density between multiple models and experiment\cite{brugman2020calcite}.
This analysis indicates that the rigid ion model\cite{armstrong2023solubility} deviates the most with respect to experiment, followed by the polarizable model\cite{raiteri2020ion}, both of which overestimate the structure of water beyond 3.5 \AA\ from the surface.
SCAN-ML is closer to experiment than the models mentioned above, and is only slightly worse than the results of AIMD with the BLYP-D3 functional\cite{brugman2020calcite}.

\subsection*{Long-range interactions via Wannier centroids}

\begin{figure}
\begin{center}
\includegraphics[width=0.6\columnwidth]{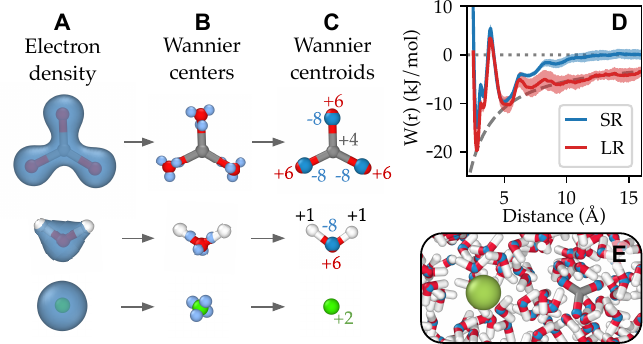}
\caption{\label{fig:Fig6} Calculation of ion pairing free energy with a ML potential with long-range interactions based on Wannier centroids. A) Isosurface of the electron density around \ce{CO3^{2-}}, water, and \ce{Ca^{2+}}. B) Wannier centers computed from DFT calculations. C) Wannier centroids and their corresponding charge. D) Ion pairing free energy for SCAN-ML (labelled SR for short-ranged) and SCAN-ML-LR (labelled LR for long-range). We have subtracted the entropic contribution to the free energy $k_B T \log (4 \pi d^2)$ and the analytic electrostatic contribution is shown using a gray dashed line. E) Snapshot of an atomic configuration extracted from the simulation used compute the ion pairing free energy for SCAN-ML-LR. Wannier centroids are shown in blue; the color for atomic species is as per Figure \ref{fig:Fig1}.}
\end{center}
\end{figure}

In the calculations above, we have neglected long-range interactions which are an essential part of the interaction of charged ions, even in the presence of a dielectric medium, such as water.
In order to incorporate these interactions, we used the framework developed by Zhang et al. \cite{Zhang21} to describe long-range interactions in insulators.
This approach is based on augmenting the short-range ML potential with the screened Coulomb interaction between electronic (negative) and nuclear (positive) spherical Gaussian charges located at Wannier centroids and at the coordinates of the nuclei, respectively.
Wannier centroids are a proxy for the center of the electronic charge distributions and can be calculated via the Wannierization of the Kohn-Sham orbitals\cite{marzari1997maximally}.
Moreover, it has been shown that the electrostatic interaction based on this description is exact up to the dipole interaction\cite{vanderbilt2018berry}.
In Figure \ref{fig:Fig6} A-C we show the electronic charge distribution around \ce{CO3^{2-}} and \ce{Ca^{2+}} ions in water, as well as for water itself, the Wannier centers, and the Wannier centroids based on the four Wannier centers around each \ce{O} or \ce{Ca} atom.
Within the pseudopotential formulation of our DFT calculations, the \ce{O}, \ce{C}, \ce{H}, and \ce{Ca} nuclei have charges +6, +4, +1, and +10.
The Wannier centroids have a charge -8, taking into account that there are four Wannier functions associated to each \ce{O} or \ce{Ca} atom, and two spin states.
For \ce{Ca} atoms we employed the effective charge +2, taking into account that the short-range ML potential captures polarization up to the cutoff.
Instead of calculating the Wannier centroids using DFT calculations at each time step, we trained a ML model for the relative position of the Wannier centroid with respect to its associated \ce{O} atom.
More information about the training of the model for the Wannier center positions is given in the Materials and Methods section.

With the new ML potential with long-range interactions, that we shall refer to as SCAN-ML-LR, we calculated the ion pairing free energy curve that we show in Figure \ref{fig:Fig6}D.
In this figure, we have removed the entropic contribution from the free energy, $G(r)$, and we shall refer to this quantity as $W(r)$ with $r$ the distance between ions (i.e. $W(r) = G(r) + k_B T \ln[4\pi r^2]$).
In the same figure, we show $W(r)$ calculated with the SCAN-ML short-range model.
In both cases, we have aligned the free energy curves such that $W(r)=0$ as $r$ tends to infinity.
In the case of the short-range model $W(r)$ goes to 0 at around twice the cutoff of the model, i.e., 12 \AA.
For SCAN-ML-LR, the interaction remains different from zero at long distances, as expected, and we have aligned the curve to the electrostatic interaction $E(r)$ of two particles with charges +2 and -2, namely, $E(r)=-4/(4 \pi \epsilon \epsilon_0 r)$, where $\epsilon_0$ is the vacuum permittivity and $\epsilon$ is the dielectric constant of the model.
An in-depth discussion of the alignment procedure for $W(r)$ and the rationale behind the removal of the entropic contribution can be found in ref.~\citenum{armstrong2023solubility}.

A key quantity that can be measured experimentally and computed from these  curves is the ion association standard free energy,
\begin{equation}
    \Delta G = -\frac{1}{\beta}\: \ln \left ( C_0 \int_0^{r^*} dr \: 4 \pi r^2 e^{- \beta W(r) } \right )
\end{equation}
where $C_0$ is the 1M concentration (1/1660 \AA$^3$), $\beta$ is the inverse temperature multiplied by Boltzmann's constant, $W(r)$ is the free energy of ion pairing properly aligned to satisfy $\lim\limits_{r \to\infty} W(r) = 0$, and $r^*=14$ \AA\ is the Bjerrum length, which defines the associated state.
$\Delta G$ for the SCAN-ML and SCAN-ML-LR are $-8.5(5)$ and $-12.5(5)$ kJ/mol, respectively.
Therefore, the long-range interactions have a non-negligible contribution to the ion association free energy, increasing it by around 50\% with respect to the short-range model.
The experimental ion association standard free energy is around $-18$ kJ/mol\cite{plummer1982solubilities,kellermeier2014straightforward}, which is around 40\% larger than the result of SCAN-ML-LR.
We note that there are several contributions to the discrepancy between SCAN-ML-LR and the experimental results.
First, the dielectric constant of water predicted by the SCAN functional is 102, which is around 25\% larger than the experimental value 78\cite{zhang2023dissolving}.
For this reason, the interaction between ions in SCAN-ML water is over-screened with respect to real water.
However, shifting $W(r)$ for SCAN-ML-LR using the experimental dielectric constant lowers the ion association free energy only by about 1 kJ/mol, not fully explaining the discrepancy.
Second, SCAN, and all current DFT functionals, have limitations in the description of many properties of aqueous systems, in addition to the dielectric constant, which may contribute to inaccurate thermodynamics of ion association. In particular, any error in the properties of liquid water will have an impact since ion pairing is driven primarily by the entropy change associated with releasing water from the solvation shells of the ions\cite{kellermeier2014straightforward}
Lastly, the experimental determination of the association constant is challenging and the reported value may have associated errors.
However, the rigid ion\cite{armstrong2023solubility} and polarizable\cite{raiteri2020ion} models predict an ion association free energy of -19 and -17 kJ/mol, respectively, both of which are consistent with experiment.

\subsection*{Conclusions}

Here, we develop a machine learning potential for calcite and calcium carbonate aqueous solutions, which we dub SCAN-ML, based on ab initio DFT calculations with the SCAN exchange and correlation functional.
A comprehensive characterization of this model has been carried out with an eye towards using it to study crystal nucleation and growth.
The results show that SCAN-ML captures diverse properties, from the details of ion pairing free energy curves to the structure of the calcite/water interface, with an accuracy comparable to or exceeding that of state-of-the-art polarizable semi-empirical force fields, and clearly superior to the best available rigid-ion models.
This is no small feat considering that decades of effort and careful work has been devoted to the fitting of semi-empirical force fields.
On the other hand, each model has a different computational cost with the SCAN DFT, SCAN-ML-LR, SCAN-ML, polarizable, and rigid ion models having a performance for MD simulations of around 0.2 ps/day, 10 ns/day, 30 ns/day, 40 ns/day, and 600 ns/day, respectively, on similar hardware and using a time step of 1 fs.
Furthermore, our calculations provide a new benchmark based on ab initio theory for many properties that are beyond the current capabilities of direct AIMD simulations.
Moreover, at variance with most semi-empirical force fields, SCAN-ML is reactive, paving the way for the study of crystallization at near neutral pH conditions where the bicarbonate to carbonate transformation plays a key role. Indeed, our results demonstrate that the initial ion pairing will be dominated by calcium bicarbonate, with the mechanism of calcium carbonate ion pair formation involving the loss of a proton from this species to hydroxide or water, as opposed to direct binding of the ions. 
Finally, our work also highlights the importance of explicitly including long-range interactions in machine learning models for accurate ion pairing thermodynamics.

\subsection*{Materials and Methods}
\label{sec:matmethods}

\paragraph{ML potential training}
ML interatomic potentials were constructed using the Deep Potential methodology developed by Zhang \textit{et al.}~\cite{Zhang18} as implemented in the \textsc{DeePMD-kit} v2.1.3 \cite{Wang18}.
We employed the smooth version of this framework\cite{Zhang18end}, which is based on deep neural networks and descriptors learned on the fly during the training process.
In our calculations, the energy was represented by a neural network with three layers and 120 neurons per layer, and the embedding matrix was represented by a three-layer neural network with sizes 25, 50 and 100.
Interactions were truncated smoothly with maximum distance $r_c=6$ \AA.
Other computational details are identical to those reported in ref.~\cite{piaggi2024first}.
The ML model with long-range interactions was trained with the Deep Potential Long Range (DPLR) framework\cite{Zhang21} using the same parameters described above for the short-range model. Additionally, the spread of the Gaussian charges was set to $0.1$ \AA$^{-1}$, and the grid size for the Fourier transform was set to 1 \AA. 

The initial training set included the following configurations: 1) $\sim$1000 configurations of liquid water and ice from ref.~\citenum{piaggi2024first} spanning the temperature range 250 to 400 K at atmospheric pressure, 2)  $\sim$1000 configurations of \ce{Ca^{2+}}, \ce{CO3^{2-}}, and an ion pair solvated by 288 water molecules obtained from MD trajectories at 300 K and atmospheric pressure based on the polarizable model of Ref.~\citenum{raiteri2020ion}, 3) ~100 configurations of the interface of calcite with water also based on the model of Ref.~\citenum{raiteri2020ion}, and 4) 100 configurations of the calcite structure in which the atomic positions and simulation box size were perturbed.
With the data described above, we trained an initial ensemble of four models, with different initialization random seeds.
The models were subsequently improved with an active learning procedure described in detail in ref.~\citenum{piaggi2024first}.

\paragraph{DFT calculations}

Plane-wave DFT calculations were performed using the \textsc{Quantum ESPRESSO} suite for electronic structure calculations v6.4.1\cite{Giannozzi09,Giannozzi17}.
The SCAN exchange and correlation functional\cite{Sun15} was evaluated with the \textsc{LIBXC} 4.3.4 library\cite{Marques12}.
We employed norm-conserving, scalar-relativistic pseudopotentials\cite{Hamann13} for \ce{Ca}, \ce{C}, \ce{O}, \ce{H} parameterized using the PBE\cite{Perdew96} functional with 10, 4, 6, and 1 valence electrons, respectively.
Kinetic energy cutoffs of 110 and 440 Ry were used for the wavefunctions and charge density.
$k$-point sampling was limited to the $\gamma$ point in all calculations after checking that this resulted in energies within 1 meV/atom of those with extensive sampling of the Brillouin zone.
The convergence absolute error for the self-consistent procedure was set to $10^{-6}$ Ry.
All other parameters were set to their default values in \textsc{Quantum ESPRESSO}.
Wannier centers were computed with Wannier90 v3.1.0\cite{Pizzi2020}.

\paragraph{Molecular dynamics}
All simulations used LAMMPS\cite{Plimpton95} patched with the DeePMD-kit\cite{Wang18} and the PLUMED enhanced sampling plugin\cite{Tribello14,Bonomi19}.
We kept the temperature constant with stochastic velocity rescaling\cite{Bussi07} using a relaxation time of 0.1 ps.
All simulations used a temperature of 330 K to compare with experiment at 300 K, allowing for the known temperature shift in properties of water in many DFT functionals, including SCAN\cite{Piaggi21,Gartner20}.
To maintain a pressure of 1 bar we employed a Parrinello-Rahman type barostat with a relaxation time of 1 ps \cite{Parrinello81}.
We used an isotropic barostat for liquids and solutions, a fully anisotropic barostat for bulk solids, and a barostat along the axis perpendicular to the surface for calcite/water interfaces.
For the calculation of static properties the mass of hydrogen was set to 2 g/mole to allow a time step of 0.5 fs, while for the calculation of dynamical properties, such as diffusion coefficients and residence times, the mass of hydrogen was set to 1 g/mole and the time step was 0.25 fs.
The masses of other elements were set to their standard values.

For calculation of the static and dynamic properties of ions, including the radial distribution functions, coordination numbers, and residence times, we used boxes with a single ion and 1576 water molecules and a total time of 10 ns.
Diffusion coefficients $D$ were calculated using Einstein's relation, which connects $D$ to the mean squared atomic displacement.
%\begin{equation}
%    D = \lim\limits_{t \to \infty} \frac{1}{6t}\langle \left | \mathbf{r}(t)-\mathbf{r}(0) \right |^2\rangle
%\end{equation}
%where $t$ is the time, $\mathbf{r}$ are the atomic coordinates, and the ensemble average $\langle\cdot\rangle$ is taken over all particles of the same type.
$D$ is known to exhibit strong finite size effects.
For this reason, we ran three different system sizes with a single ion in 197, 788, or 1576 water molecules, and  extrapolated $D$ to the infinite system size limit.
The lattice constants of calcite, aragonite, and monohydrocalcite were computed over a 1-ns simulation for systems of 2160, 1280, and 864 atoms.
We studied the calcite/water interface using a system of 10584 atoms and box sides 40x48x60 \AA, and for a total time of 10 ns.
%This simulation was used to compute all quantities reported in Figure \ref{fig:Fig5} and the residence time reported in Table \ref{tab:Table1}.

\paragraph{Enhanced sampling}

The ion pairing free energy curves were computed using On-the-fly Probability Enhanced Sampling (OPES)\cite{Invernizzi20rethinking}, which is an evolution of the widely-used technique Metadynamics\cite{Laio02}.
We used as a CV the distance $d$ between the \ce{Ca} and \ce{C} atoms.
The bias update frequency for the OPES simulations was 500 steps, and the barrier for maximum free energy exploration was set to 30 kJ/mol.
We performed four independent simulations, each starting from different initial velocities compatible with the Maxwell-Boltzmann distribution, and used systems with one ion pair and 197, 1576, and 5319 water molecules in cubic boxes with side lengths 18, 36, and 53 \AA, respectively, to check that our results do not depend on the system size (free energy curves for different system sizes can be found in Figure S1 in the Supplementary Material).
A harmonic bias potential upper wall was introduced at 9, 15, and 20 \AA\ for the systems with box sides of 18, 36, and 53 \AA, respectively, with a force constant of $2\times10^3$ kJ/mol/nm.
This wall prevented the exploration of distances beyond around half the box length.
In Figure \ref{fig:Fig3}B and \ref{fig:Fig6}D we report the results of four 20-ns-long simulations with the intermediate system size, namely, one ion pair in 1576 water molecules.

Within $\sim$ 1 ns, each OPES simulation converges to a quasi-static bias potential $V$ and then explores reversibly the whole free energy landscape for the remainder of the simulation.
Afterwards, we can compute the free energy surface (FES) via umbrella-sampling-type re-weighting,
\begin{equation}
    F(\mathbf{s})=-\frac{1}{\beta} \ln \left ( \frac{\langle \delta(\mathbf{s}-\mathbf{s}(\mathbf{R})) e^{\beta V(\mathbf{R})} \rangle_V}{\langle e^{\beta V(\mathbf{R})} \rangle_V} \right)
\end{equation}
where $\mathbf{R}$ are the atomic coordinates, $\mathbf{s}$ are the CVs for re-weighting and $\langle \cdot \rangle_V$ is an  average over $\mathbf{R}$ in the biased ensemble.
Note that the CVs $\mathbf{s}$ chosen for re-weighting are not necessarily the same as those used to construct the bias potential $V$.
Indeed, while only $d$ was biased to construct Figure \ref{fig:Fig3}, we computed $F(\mathbf{s})$ using $d$ and the Ca-O$_\mathrm{w}$ coordination number as CVs.
The Ca-O$_\mathrm{w}$ coordination number was computed smoothly via a cubic switching function, so that atoms beyond $D_\mathrm{max}=3.75$ \AA\ had zero weight, those within $D_0=2.5$ \AA\ had a weight of one, and atoms in between have intermediate weights (see ref.~\cite{Tribello14} for further details).

For the simulation of the carbonate to bicarbonate transformation (Figure \ref{fig:Fig4}) we used $d$ and the O$_\mathrm{c}$-H coordination number as CVs to construct the bias potential.
To obtain a continuous and differentiable O$_\mathrm{c}$-H coordination number we defined a cubic switching function with $D_0=0.95$ \AA\ and $D_\mathrm{max}=1.6$ \AA.
We simulated a system with a single ion pair in 197 water molecules for a total time of 50 ns.
During this simulation, there were six transitions between carbonate and bicarbonate, and for each of these species the variable $d$ was thoroughly explored, indicating the reliability of the calculations.

\subsection*{Data availability}
The training data, ML models, and all input files to reproduce the simulations are available for download at \url{https://doi.org/10.5281/zenodo.13834650}, and in PLUMED NEST\cite{Bonomi19}, the public
repository of the PLUMED consortium, as \href{https://www.plumed-nest.org/eggs/24/021/}{plumID:24.021}.

\subsection*{Acknowledgements}

P.M.P. acknowledges funding from the Marie Skłodowska-Curie Cofund Programme of European Commission project H2020-MSCA-COFUND-2020-101034228-WOLFRAM2. We acknowledge use of multiple high-performance computing facilities; 1) the National Energy Research Scientific Computing Center (NERSC), a Department of Energy Office of Science User Facility using NERSC award BES-ERCAP-0029357), 2) RES resources provided by BSC in MareNostrum to RES-FI-2024-2-0026, and 3) Princeton Research Computing resources, Princeton University. P.R. and J.D.G. thank the Australian Research Council for funding, and the Pawsey Supercomputing Centre and National Computational Infrastructure for computing resources.
% Bibliography
%\bibliographystyle{rsc}
%\bibliography{mybib}

\providecommand*{\mcitethebibliography}{\thebibliography}
\csname @ifundefined\endcsname{endmcitethebibliography}
{\let\endmcitethebibliography\endthebibliography}{}

% If your first paragraph (i.e. with the \dropcap) contains a list environment (quote, quotation, theorem, definition, enumerate, itemize...), the line after the list may have some extra indentation. If this is the case, add \parshape=0 to the end of the list environment.

\end{document}